\documentclass[runningheads]{llncs}
\usepackage{algorithm}
\usepackage{algorithmic}
\usepackage{amsmath}
\usepackage{amsfonts}
\usepackage{xcolor}
\usepackage{multirow}
\usepackage{comment}
\usepackage{verbatim}
\usepackage{adjustbox}
\usepackage{graphicx}
\usepackage{lscape}
\usepackage{tabularx}
\usepackage{booktabs}
\usepackage{hyperref}
\begin{document}
\title{Patch-based Generative Adversarial Network Towards Retinal Vessel Segmentation} 
\titlerunning{Conditional Patch-based GAN for Retinal Vessel Segmentation}
%
\author{Waseem Abbas\inst{1}\orcidID{0000-0003-1978-1161} \and
Muhammad Haroon Shakeel\inst{2}\orcidID{0000-0001-6237-3388} \and
Numan Khurshid\inst{2}\orcidID{0000-0002-8263-4781} \and Murtaza Taj\inst{2}\orcidID{0000-0003-2353-4462}}
\authorrunning{W. Abbas et al.}
%
\institute{Cloud Application Solutions Division, \\Mentor, A Siemens Business, Lahore, Pakistan \\
\email{muhammad\_waseem@mentor.com}
\and
Department of Computer Science, Syed Babar Ali School of Science and Engineering, Lahore University of Management Sciences (LUMS), Lahore, Pakistan
\\
\email{\{15030040,15060051,murtaza.taj\}@lums.edu.pk}}
\maketitle              

\begin{abstract}
Retinal blood vessels are considered to be the reliable diagnostic biomarkers of ophthalmologic and diabetic retinopathy. Monitoring and diagnosis totally depends on expert analysis of both thin and thick retinal vessels which has recently been carried out by various artificial intelligent techniques. Existing deep learning methods attempt to segment retinal vessels using a unified loss function optimized for both thin and thick vessels with equal importance. Due to variable thickness, biased distribution, and difference in spatial features of thin and thick vessels, unified loss function are more influential towards identification of thick vessels resulting in weak segmentation. To address this problem, a conditional patch-based generative adversarial network is proposed which utilizes a generator network and a patch-based discriminator network conditioned on the sample data with an additional loss function to learn both thin and thick vessels. Experiments are conducted on publicly available STARE and DRIVE datasets which show that the proposed model outperforms the state-of-the-art methods.

\end{abstract}

\section{Introduction}

Deep learning has influenced image analysis in various important application areas including remote-sensing, autonomous vehicles, and specially medical~\cite{nazir2019tiny,abbas2019adaptively}. Usually, diagnostic analysis and treatment which covers medical disorders, diabetic retinopathy, and glaucoma have been carried on retinal vessels of opthalmologic fundus images~\cite{patton2006retinal}. Morphological attributes and retinal vascular structures including vascular tree patterns, vessels thickness, color, density, crookedness, and relative angles are the key features used in the diagnostic process by ophthalmologists~\cite{fraz2012blood}. Thickness and visibility of retinal vessels are the core attributes for diabetic retinopathy analysis~\cite{abramoff2010retinal}. However, conventional and manual approaches for vascular analysis are time-consuming and prone to human error. Therefore, computer-assisted detection of retinal vessels is inevitable to segment out the retinal vessels from fundus images and mainly categorized into three approaches: \textit{unsupervised learning}, \textit{supervised learning}, and \textit{deep learning}.

Unsupervised learning approaches extracts vessel pattern without class labels, mainly thorough filter-based approach. Image filters, such as Gaussian blur, are utilized to enhance vascular features. Zhang \textit{et al.} combined the matched filter and first-order derivative of Gaussian filter to extract retinal vessel features~\cite{zhang2010retinal}.  Similarly, Fraz \textit{et al.} applied the same approach in four directions along with multi-directional morphological top-hat operator to detect retinal vessels~\cite{fraz2012approach}.  Yin \textit{et al.} worked on orientation invariant approach and used Fourier transform to extract energy maps, consequently detecting retinal vessels using an orientation-aware detector~\cite{yin2015vessel}. A similar approach of using Wavelet transform to map images into a 3D lifted-domain was proposed in ~\cite{zhang2016robust}. They utilized the Gaussian filter to detect retinal vessels.

Supervised approaches intend to assign a class label to each pixel of the image. These approaches can be based on either traditional \textit{machine learning} or \textit{deep learning}. The former utilizes classifiers that learn decision boundaries on handcrafted features e.g. Support Vector Machine (SVM) and K-nearest neighbor classifier (KNN).

However, more recently, Convolutional Neural Networks (CNNs) got popularity for segmentation problems since they learn the features directly from the input data. In the case of retinal vessel segmentation, CNNs surpass traditional handcrafted approaches~\cite{melinvsvcak2015retinal}. However, the problem of blurred output images and false positives around indistinct tiny vessel branches is persistent in these methods. The main reason behind this limitation is use of a unified loss function in a pixel-wise manner to segment both thin and thick vessels. This led to blurred thin vessels resulting in non acceptable segmentation maps during binarization of generated probability maps. To address these issues a novel approach for retinal vessel segmentation using Generative Adversarial Networks (GAN) has been proposed in this research. Basically, a patch-based discriminator is utilized to learn inconsistencies and sharp edges of high-resolution blood vessels. Additionally, a loss term is integrated within the main objective function to learn low-frequency edges.  We show that the proposed method is able to effectively segment out thick and thin vascular pixels form non-vascular pixels on two benchmark datasets namely DRIVE~\cite{niemeijer2004drive} and STARE~\cite{hoover2000locating}. Our results show a significant boost in the performance as compare to the state-of-the-art methods.

\section{Methodology}

\begin{figure*}[t]
\vskip 0.2in
\begin{center}
\centerline{\includegraphics[width=1\textwidth]{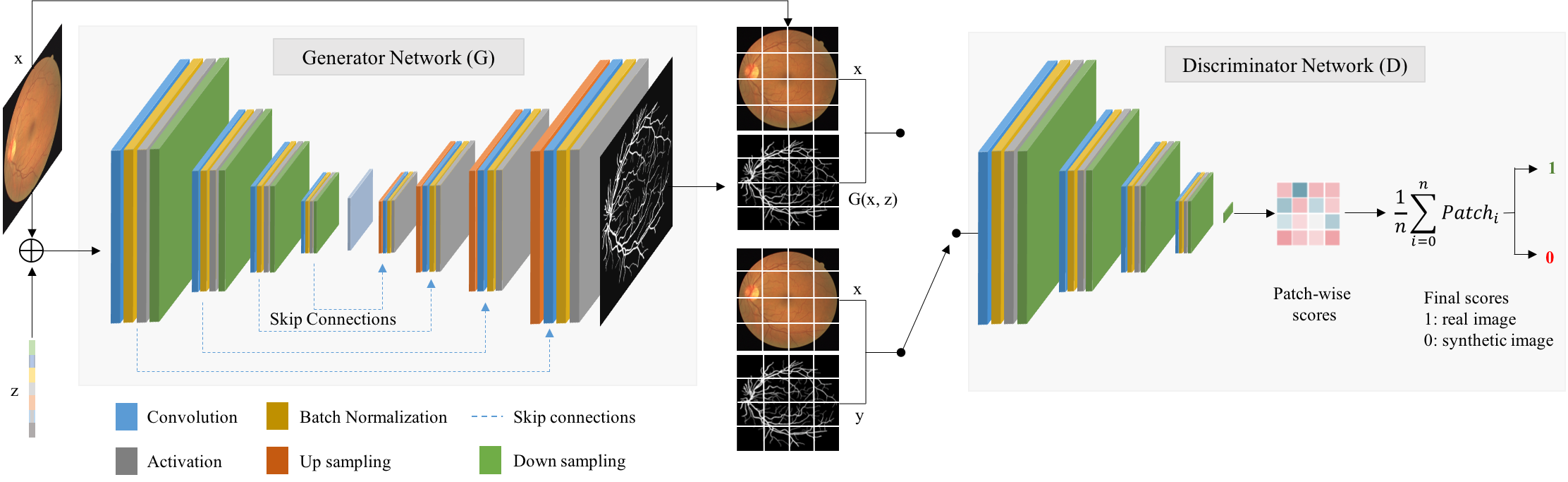}}
\caption{The proposed Conditional Patch-based Generative Adversarial Network.}
\label{model}
\end{center}
\vskip -0.4in
\end{figure*}

Adversarial learning approach combines generator and discriminator networks in such a way that conditional input provides a head start to the overall learning process. 
Generator network $G$ expects noise and conditional input sample and learn to generate the synthetic retinal map. Discriminator network $D$ takes two sets of input: conditional input and a generated synthetic map (fake sample) and conditional input and actual segmentation map (/real sample/ground truth). Segmentation maps generated from generator network are divided into rectangular patches and discriminator tries to discriminate each patch. This patch based discriminator is deployed to discriminate between actual or synthetically generated segmentation maps as shown in Fig.~\ref{model}.

\subsection{Objective Function}

In adversarial learning, generator network $G$ tries to map input conditional sample $x^{w\times h}$ and noise vector to its corresponding segmentation map $y^{w\times h}$ in encoder/decoder arrangement such that the difference between $y^{w\times h}$ and synthetic vessel map $G(x, z)$ is reduced ($w$:width and $h$:height of image). Discriminator network takes two pairs $\{x,y\}$ and \{$x,G(x, z)\}$ as input and predicts the score between 0 or 1 as $\{0,1\}^n$ where $n$ is a hyperparameter of the model and represents the total number of patches fed to the discriminator $D$. $n$ could be selected between $0$ and the total number of pixels of image. The objective function of the proposed model can be formulated as:

\begin{equation} \label{ggan}
\begin{aligned}
J = \arg \operatorname*{min}_G \operatorname*{max}_D  {\mathbb{E}}_{x,y}\big(-log(D(x,y)\big)+{\mathbb{E}}_{x,z}\big(-log(1-D(x,G(x,z)))\big) \\ + \lambda \mathcal{L}_{L_1}(G)
\end{aligned}
\end{equation}
where $\mathbb{E}_{x,z}$ and $\mathbb{E}_{x,y}$ are loss functions for generator $G$ and discriminator $D$ respectively. To handle the problem of blurred outputs caused by $L_2$ norm reported in literature, we integrated $L_1$ norm in the main objective function to capture low and high-frequency components of the fundus retinal images. $L_1$ norm can be formulated as:

\begin{equation} \label{ggan2}
\begin{aligned}
\mathcal{L}_{L_1}(G) = {\mathbb{E}}_{x,y,z} \Big[||y-G(x,z)||_1\Big]
\end{aligned}
\end{equation}

where $\lambda$ is a hyperparameter. The noise vector $z$ is selected from a normal distribution to ensure that the generator learns a random sample at each training step. This also prevents the local minima problems.

\subsection{Patch-based Discrimination}

To detect high-frequency components and thin vessels, patch-based discriminator network is proposed which penalize each patch and discriminate real or generated synthesized segmentation maps. This kind of approach treats each individual rectangular patch as a stand-alone image and results in the probability map on each patch. The final result against an image is obtained by averaging all patch based results. Patch based discriminator processes input as a Markov random field by assuming independence among all patches. The small size of the patch allows fast convergence of network and results in high-resolution segmentation maps.

\subsection{Model Architecture}

To extract low level features, proposed model followed inspiration form UNet~\cite{ronneberger2015u} and SegNet~\cite{badrinarayanan2017segnet}. The generative network is comprised of encoder and decoder networks. Encoder network extracts the hidden features and reduces the size of the input image whereas the decoder network reconstructs and up-sample at each stage till the last layer. Each input image passes through the entire generator network and skip connections between encoder and decoder network acts as bridge to semantic gap between the feature maps of the encoder and decoder prior to fusion.

\subsection{Hyperparameter Tuning}

The proposed model is trained in a manner such that a the generator network is trained and generates random output regardless of input sample and tries to learn the patterns of input as per given ground truths. Meanwhile, the discriminator network tries to discriminate the generated outputs of generator network and the ground truths. Further, discriminator learns to utilize the conditional samples ($x$) to learn the pattern of blood vessels in fundoscopic images. Stochastic gradient decent with Adam optimizer is used to train the generator network. All the hyperparameters used in training of the proposed model are selected empirically, which include trade-off coefficient ($\lambda = 10$), learning rate (lr = 0.002), learning rate decaying factor ($\eta$ = 0.75) and momentum (m = 0.002).

\subsection{Datasets and Evaluation Metrices}

The proposed model is trained and evaluated on two publicly available datasets include DRIVE~\cite{niemeijer2004drive}, STARE~\cite{hoover2000locating}. We followed the same evaluation metrics and protocols to conduct a fair evaluation with the reported state-of-the-art methods. Accuracy ($Acc$), sensitivity ($Se$) and specificity ($Sp$) are used as a benchmark for quantitative evaluation and area under the receiving operating curve ($AUC$) is used for the qualitative evaluation.

\section{Results and Discussion}

\begin{table*}[t]
\centering
\caption{Performance comparison of the proposed model with sate-of-the-art methods.}
\label{comp}
\begin{center}
\begin{small}
\begin{sc}
\vskip -0.25in
\resizebox{1\textwidth}{!}{%
\begin{tabular}{lcc|cccc|cccc}
\toprule
\multirow{2}{*}{{Scheme}} & \multirow{2}{*}{{Methods}} &
\multirow{2}{*}{{Year}} & \multicolumn{4}{c}{{DRIVE}} & \multicolumn{4}{c}{{STARE}}
\\ \cmidrule{4-11} 
& & & {$Acc$} & {$Sp$}& {$Se$}& {$AUC$}& {$Acc$} & {$Sp$}& {$Se$}& {$AUC$}\\ \midrule

& Human Observer & & 0.9472 & 0.9724 & 0.7760 & - & 0.9349 & 0.9384 & 0.8952 & -  \\ \midrule

& zhang~\cite{zhang2010retinal} & 2010 & 0.9382 & 0.9724 & 0.7120 & - & 0.9484 & 0.9753 & 0.7177 & - \\

& Fraz~\cite{fraz2012approach} & 2012 & 0.9430 & 0.9759 & 0.7152 & - & 0.9442 & 0.9686 & 0.7311 & - \\

\multirow{2}{*}{{Unsupervised}}& Roychowdhuray~\cite{roychowdhury2015iterative} & 2015 & 0.9494 & 0.9782 & 0.7395 & 0.9672 & 0.9560 & 0.9842 & 0.7317 & 0.9673 \\

& Azzopardi~\cite{azzopardi2015trainable} & 2015 & 0.9442 & 0.9704 & 0.7655 & 0.9614 & 0.9497 & 0.9701 & 0.7716 & 0.9563 \\

& Yin~\cite{yin2015vessel} & 2015 & 0.9403 & 0.9790 & 0.7246 & - & 0.9325 & 0.9419 & \textbf{0.8541} & - \\

& Zhang~\cite{zhang2016robust} & 2016 & 0.9476 & 0.9725 & 0.7743 & 0.9636 & 0.9554 & 0.9758 & 0.7791 & 0.9748  \\ \midrule

\multirow{5}{*}{{Supervised}} & You~\cite{you2011segmentation} & 2011 & 0.9434 & 0.9751 & 0.7410 & - & 0.9497 & 0.9756 & 0.7260 & - \\

& Marin~\cite{marin2011new} & 2011 & 0.9452 & 0.9801 & 0.7067 & 0.9588 & 0.9526 & 0.9819 & 0.6944 & 0.9769 \\

& Fraz~\cite{fraz2012ensemble} & 2012 & 0.9480 & 0.9807 & 0.7406 & 0.9747 & 0.9534 & 0.9763 & 0.7548 & 0.9768 \\

& Orlando~\cite{orlando2017discriminatively} & 2017 & - & 0.9684 & \textbf{0.7897} & - & - & 0.9738 & 0.7680 & - \\

& Dasgupta~\cite{dasgupta2017fully} & 2017 & 0.9533 & 0.9801 & 0.7691 & 0.9744 &  - & - & - & - \\
\midrule

\multirow{6}{*}{{Deep Learning}} & Melin~\cite{melinvsvcak2015retinal} & 2015 & 0.9466 & 0.9785 & 0.7276 & 0.9749 & - & - & - & - \\

& Li~\cite{li2016cross} & 2016 & 0.9527 & 0.9816 & 0.7569 & 0.9738 & 0.9628 & 0.9844 & 0.7726 & 0.9879 \\

& Liskowski~\cite{liskowski2016segmenting} & 2016 & 0.9515 & 0.9806 & 0.7520 & 0.9710 & \textbf{0.9696} & 0.9866 & 0.8145 & 0.9880 \\

& Fu~\cite{fu2016retinal} & 2016 & 0.9523 & - & 0.7603 & - & 0.9585 & - & 0.7412 & - \\

& Zengqiang~\cite{8476171} & 2018 & 0.9538 & 0.9820 & 0.7631 & 0.9750 & 0.9636 & {0.9857} & 0.7735 & 0.9833 \\

& \textbf{Proposed} & 2019 & \textbf{0.9562} & \textbf{0.9824} & 0.7746 & \textbf{0.9753} & {0.9647} & \textbf{0.9869} & 0.7940 & \textbf{0.9885} \\ \bottomrule

\end{tabular}
}

\end{sc}
\end{small}
\end{center}
\vskip -0.25in

\end{table*}

Evaluation of the proposed model is conducted on DRIVE and STARE datasets and categorized into unsupervised, supervised and deep learning schemes as summarized in Table~\ref{comp}. $Acc$, $Se$, $Sp$ and $AUC$ values are mentioned against the proposed method and reported in the literature.

In unsupervised techniques, the most recent findings were reported in~\cite{zhang2016robust} where researchers used left invariant rotating derivative to get enhanced retinal vessels and obtained binary segmentation using thresholding. Their method outperformed the previous unsupervised techniques on DRIVE and STARE datasets~\cite{azzopardi2015trainable,fraz2012ensemble,roychowdhury2015iterative,yin2015vessel,zhang2010retinal}. In supervised learning techniques, ~\cite{dasgupta2017fully} achieved best results on DRIVE dataset by deploying a combination of convolutional neural network and structured predictions. The second best method \cite{fraz2012approach} in supervised learning used conditional random field model with a fully connected method and achieved comparable performance on DRIVE and STARE datasets as compared to other supervised learning schemes~\cite{fraz2012ensemble,marin2011new,you2011segmentation}.

\begin{figure}[!t]
\begin{center}
\centerline{\includegraphics[width=0.9\textwidth]{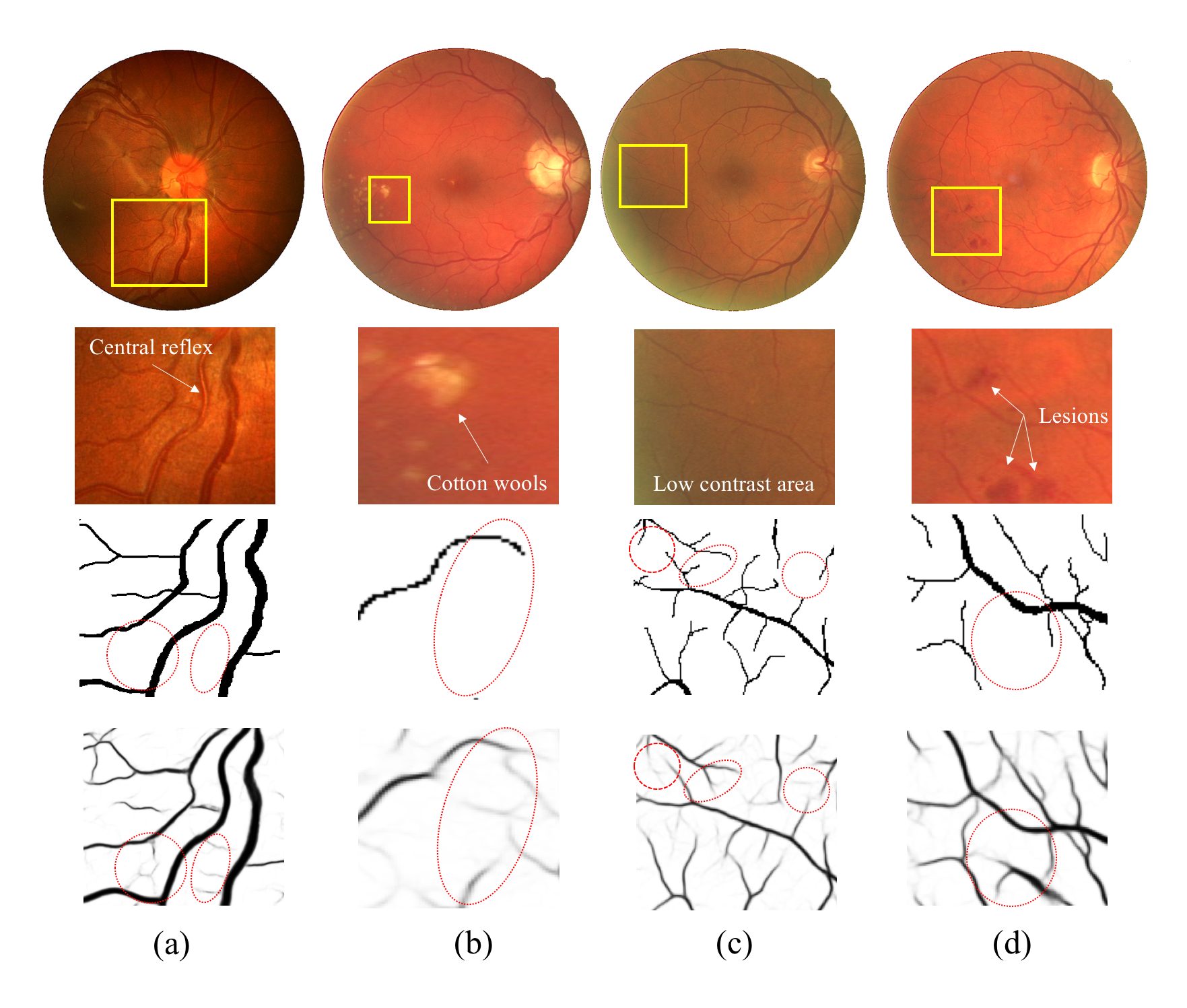}}
\caption{Exemplar results of the proposed model on challenging cases: (a): central reflex vessels, (b): cotton wools, (c): low contrast, (d) lesions. From top to bottom: input fundus image, enlarged target patch of fundus image, corresponding manual annotation and the predicted probability maps.}
\label{cases}
\end{center}
\vskip -0.5in
\end{figure}

For the DRIVE dataset, the proposed model achieves 0.9562, 0.9824, 0.7746 and 0.9753 for $Acc$, $Sp$, $Se$ and $AUC$ respectively, where the model achieves better results for $Acc$, $Sp$ and $AUC$ as compared to the all current state-of-the-art unsupervised, supervised and deep learning techniques. However Orlando~\cite{orlando2017discriminatively} outperforms all the methods in terms of $Se$ as shown in Table~\ref{comp}, the only $Se$ norm is not conclusive. In contrast, the performance of the proposed model is much better than all the compared methods.

On the STARE fundoscopic image dataset, the proposed model achieves 0.9647, 0.9862, 0.7940 and 0.9885 for $Acc$, $Sp$, $Se$ and $AUC$ respectively. In terms of $Sp$ and $AUC$, the proposed model outperforms all the compared techniques. However, Yin~\cite{yin2015vessel} reported the best performance in terms of $Se$ by achieving 0.0601 more sensitivity but obtains 0.030 lesser specificities. Similarly, Zengqiang~\cite{8476171} achieves better results in terms of $Acc$ by achieving $0.9696$ accuracy but lags in other evaluation benchmarks as compared to the proposed method.

Presence of lesions and cotton wools in fundoscopic images mainly affect the local features and the thick vessels. Other challenges are the presence of central reflex vessels and low contrasts. To address four types of challenges (central reflex vessels, cotton wools, low contrast, and lesions) in segmentation of fundoscopic images, the proposed model is able to segment out the retinal vessels in these challenging scenarios. By integrating $L_1$ norm in the main objective function, the generator network is able to detect low contrast and thin retinal vessels as shown in Fig.~\ref{cases}. The generator network learns the low contrast vessels whereas the discriminator network forces the model to learn the non-vessel pixels too by predicting a zero score. In this way, the entire model learns the structure and appearance of vessels simultaneously and the model is able to address the central reflex vessel problem. Patch based discriminator network allows the model to capture thin vessels in the presence of lesions and cotton wools. In summary, the proposed generative adversarial network can effectively address the main challenging cases by learning generator and discriminator network alternatively and integrating a custom loss term.

\section{Conclusion}

A novel generative adversarial network based deep learning model has been proposed, that can potentially address segmentation of retinal blood vessels in fundoscopic images. Training the generator network to learn small transitions in thin vessels and allowing the patch based discriminator to discriminate vascular and non-vascular pixels. Results on publicly available datasets showed that the proposed model is competitive with current state-of-the-art techniques. Averaging the patch based results over small patches of fundoscopic image and integration of additional loss term into the main objective function leverage and enhances the effectiveness of the proposed model. The model has the potential to probe the different patch sizes so that the influence of patch-based discriminator on segmentation performance can be better analyzed.

\bibliographystyle{splncs04}
\bibliography{mybibfile}

\begin{thebibliography}{10}
\providecommand{\url}[1]{\texttt{#1}}
\providecommand{\urlprefix}{URL }
\providecommand{\doi}[1]{https://doi.org/#1}

\bibitem{abbas2019adaptively}
Abbas, W., Taj, M.: Adaptively weighted multi-task learning using inverse
  validation loss. In: IEEE International Conference on Acoustics, Speech and
  Signal Processing. pp. 1408--1412 (2019)

\bibitem{abramoff2010retinal}
Abr{\`a}moff, M.D., Garvin, M.K., Sonka, M.: Retinal imaging and image
  analysis. IEEE Reviews in Biomedical Engineering  \textbf{3},  169--208
  (2010)

\bibitem{azzopardi2015trainable}
Azzopardi, G., Strisciuglio, N., Vento, M., Petkov, N.: Trainable cosfire
  filters for vessel delineation with application to retinal images. Medical
  Image Analysis  \textbf{19}(1),  46--57 (2015)

\bibitem{badrinarayanan2017segnet}
Badrinarayanan, V., Kendall, A., Cipolla, R.: Segnet: A deep convolutional
  encoder-decoder architecture for image segmentation. IEEE Transactions on
  Pattern Analysis and Machine Intelligence  \textbf{39}(12),  2481--2495
  (2017)

\bibitem{dasgupta2017fully}
Dasgupta, A., Singh, S.: A fully convolutional neural network based structured
  prediction approach towards the retinal vessel segmentation. In: IEEE
  International Symposium on Biomedical Imaging. pp. 248--251 (2017)

\bibitem{fraz2012approach}
Fraz, M.M., Barman, S.A., Remagnino, P., Hoppe, A., Basit, A., Uyyanonvara, B.,
  Rudnicka, A.R., Owen, C.G.: An approach to localize the retinal blood vessels
  using bit planes and centerline detection. Computer Methods and Programs in
  Biomedicine  \textbf{108}(2),  600--616 (2012)

\bibitem{fraz2012blood}
Fraz, M.M., Remagnino, P., Hoppe, A., Uyyanonvara, B., Rudnicka, A.R., Owen,
  C.G., Barman, S.A.: Blood vessel segmentation methodologies in retinal
  images--a survey. Computer Methods and Programs in Biomedicine
  \textbf{108}(1),  407--433 (2012)

\bibitem{fraz2012ensemble}
Fraz, M.M., Remagnino, P., Hoppe, A., Uyyanonvara, B., Rudnicka, A.R., Owen,
  C.G., Barman, S.A.: An ensemble classification-based approach applied to
  retinal blood vessel segmentation. IEEE Transactions on Biomedical
  Engineering  \textbf{59}(9),  2538--2548 (2012)

\bibitem{fu2016retinal}
Fu, H., Xu, Y., Wong, D.W.K., Liu, J.: Retinal vessel segmentation via deep
  learning network and fully-connected conditional random fields. In: IEEE
  International Symposium on Biomedical Imaging. pp. 698--701 (2016)

\bibitem{hoover2000locating}
Hoover, A., Kouznetsova, V., Goldbaum, M.: Locating blood vessels in retinal
  images by piecewise threshold probing of a matched filter response. IEEE
  Transactions on Medical imaging  \textbf{19}(3),  203--210 (2000)

\bibitem{li2016cross}
Li, Q., Feng, B., Xie, L., Liang, P., Zhang, H., Wang, T.: A cross-modality
  learning approach for vessel segmentation in retinal images. IEEE Transaction
  on Medical Imaging  \textbf{35}(1),  109--118 (2016)

\bibitem{liskowski2016segmenting}
Liskowski, P., Krawiec, K.: Segmenting retinal blood vessels with deep neural
  networks. IEEE Transactions on Medical Imaging  \textbf{35}(11),  2369--2380
  (2016)

\bibitem{marin2011new}
Mar{\'\i}n, D., Aquino, A., Geg{\'u}ndez-Arias, M.E., Bravo, J.M.: A new
  supervised method for blood vessel segmentation in retinal images by using
  gray-level and moment invariants-based features. IEEE Transactions on Medical
  Imaging  \textbf{30}(1), ~146 (2011)

\bibitem{melinvsvcak2015retinal}
Melin{\v{s}}{\v{c}}ak, M., Prenta{\v{s}}i{\'c}, P., Lon{\v{c}}ari{\'c}, S.:
  Retinal vessel segmentation using deep neural networks. In: International
  Conference on Computer Vision Theory and Applications) (2015)

\bibitem{nazir2019tiny}
Nazir, U., Khurshid, N., Ahmed~Bhimra, M., Taj, M.: Tiny-inception-resnet-v2:
  Using deep learning for eliminating bonded labors of brick kilns in south
  asia. In: Proceedings of the IEEE Conference on Computer Vision and Pattern
  Recognition Workshops. pp. 39--43 (2019)

\bibitem{niemeijer2004drive}
Niemeijer, M., Staal, J., Ginneken, B., Loog, M., Abramoff, M.: Drive: digital
  retinal images for vessel extraction. Methods for Evaluating Segmentation and
  Indexing Techniques Dedicated to Retinal Ophthalmology  (2004)

\bibitem{orlando2017discriminatively}
Orlando, J.I., Prokofyeva, E., Blaschko, M.B.: A discriminatively trained fully
  connected conditional random field model for blood vessel segmentation in
  fundus images. IEEE Transactions on Biomedical Engineering  \textbf{64}(1),
  16--27 (2017)

\bibitem{patton2006retinal}
Patton, N., Aslam, T.M., MacGillivray, T., Deary, I.J., Dhillon, B., Eikelboom,
  R.H., Yogesan, K., Constable, I.J.: Retinal image analysis: concepts,
  applications and potential. Progress in Retinal and Eye Research
  \textbf{25}(1),  99--127 (2006)

\bibitem{ronneberger2015u}
Ronneberger, O., Fischer, P., Brox, T.: U-net: Convolutional networks for
  biomedical image segmentation. In: International Conference on Medical Image
  Computing and Computer-assisted Intervention. pp. 234--241 (2015)

\bibitem{roychowdhury2015iterative}
Roychowdhury, S., Koozekanani, D.D., Parhi, K.K.: Iterative vessel segmentation
  of fundus images. IEEE Transactions on Biomedical Engineering
  \textbf{62}(7),  1738--1749 (2015)

\bibitem{8476171}
Yan, Z., Yang, X., Cheng, K.T.: A three-stage deep learning model for accurate
  retinal vessel segmentation. IEEE Journal of Biomedical and Health
  Informatics pp. 1427--1436 (2018)

\bibitem{yin2015vessel}
Yin, B., Li, H., Sheng, B., Hou, X., Chen, Y., Wu, W., Li, P., Shen, R., Bao,
  Y., Jia, W.: Vessel extraction from non-fluorescein fundus images using
  orientation-aware detector. Medical Image Analysis  \textbf{26}(1),  232--242
  (2015)

\bibitem{you2011segmentation}
You, X., Peng, Q., Yuan, Y., Cheung, Y.m., Lei, J.: Segmentation of retinal
  blood vessels using the radial projection and semi-supervised approach.
  Pattern Recognition  \textbf{44}(10-11),  2314--2324 (2011)

\bibitem{zhang2010retinal}
Zhang, B., Zhang, L., Zhang, L., Karray, F.: Retinal vessel extraction by
  matched filter with first-order derivative of gaussian. Computers in Biology
  and Medicine  \textbf{40}(4),  438--445 (2010)

\bibitem{zhang2016robust}
Zhang, J., Dashtbozorg, B., Bekkers, E., Pluim, J.P., Duits, R., ter
  Haar~Romeny, B.M.: Robust retinal vessel segmentation via locally adaptive
  derivative frames in orientation scores. IEEE transactions on Medical Imaging
   \textbf{35}(12),  2631--2644 (2016)

\end{thebibliography}

\end{document}